\documentclass[fleqn,usenatbib]{mnras}

\usepackage{newtxtext,newtxmath}

\usepackage[T1]{fontenc}

\DeclareRobustCommand{\VAN}[3]{#2}
\let\VANthebibliography\thebibliography
\def\thebibliography{\DeclareRobustCommand{\VAN}[3]{##3}\VANthebibliography}

\usepackage{graphicx}	
\usepackage{amsmath}	
\usepackage{hyperref}

\title[Short title, max. 45 characters]{Impact of Jet Density on Intracluster Medium Heating in Self-Regulated AGN Feedback Simulations}

\author[]{
Tzu-Wei Tsai,$^{1}$
Hsiang-Yi Karen Yang$^{2,3}$
\\
$^{1}$Department of Physics, National Taiwan University, Taipei 10617, Taiwan\\
$^{2}$Institute of Astronomy, National Tsing Hua University, Hsinchu 30013, Taiwan\\
$^{3}$Physics Division, National Center for Theoretical Sciences, Taipei 10617, Taiwan
}

\date{Accepted XXX. Received YYY; in original form ZZZ}

\pubyear{\the\year{}}

\begin{document}
\label{firstpage}
\pagerange{\pageref{firstpage}--\pageref{lastpage}}
\maketitle

\begin{abstract}
Active galactic nucleus (AGNs) feedback is widely accepted as the key mechanism to suppress cooling flows in galaxy clusters. However, the dependence of heating efficiency on jet properties is not fully understood. In this work, we present three-dimensional hydrodynamic simulations of a Perseus-like cluster, including both single-jet and self-regulated models, to investigate how jet density affects bubble evolution and the thermal balance of the intracluster medium. Our results confirm previous findings that lighter jets inflate more spherical bubbles and are more easily deflected by cold gas, enabling isotropic energy deposition throughout the cluster core. However, despite their broader spatial impact, lighter jets display lower overall heating efficiency, requiring higher average jet power to maintain self-regulation compared to heavier jets. We also find that the distribution and amount of cold gas significantly influence the effectiveness of jet heating. These results highlight jet density as a critical parameter in AGN feedback and emphasize the need to incorporate additional physical processes such as magnetic fields, viscosity, and cosmic rays in future studies for realistic comparisons with observations.
\end{abstract}




\section{Introduction}
A galaxy cluster with a short central cooling time is known as a cool-core (CC) cluster. These clusters exhibit high central gas density ($\gtrsim 10^{-2} \text{ cm}^{-3}$) and low central entropy ($\lesssim 50 \text{ keV cm}^2$), conditions that are expected to lead to significant gas inflows. This flow is predicted by the "cooling-flow" model \citep{Fabian_1994} and could induce the intensive star formation in the central brightest cluster galaxy. However, such inflows and  the predicted high star formation rates (SFRs) have not been observed \citep{PETERSON20061}. This is the "cooling-flow problem".\\

Active galactic nucleus (AGN) feedback at the cluster center is now widely accepted as a key solution to the cooling flow problem (see reviews by \citet{McNamara_2012, Bourne_2023}). In simulations, insufficient AGN feedback can result in unrealistically high SFRs compared to observational data \citep{Li_2015}. Meanwhile, X-ray observations have provided compelling evidence for AGN feedback and its associated heating effects in galaxy clusters. \citep{Fabian_2012, Dunn_2008}. This feedback expels gas from the core and heats the intracluster medium (ICM), counteracting radiative cooling and thereby reducing the SFRs as observed. The outflows are intermittent and are triggered by rapid variations in accretion via the process of chaotic cold accretion \citep{Gaspari_2012}. After the cluster cools again, another cycle of feedback is initialed. With sustained AGN feedback, the cluster can become self-regulated, where heating and cooling are globally balanced \citep{Gaspari_2011,Li_2015,Prasad_2015,Karen_Yang_2016,Li_2017,Beckmann_2019,Ehlert_2022}. \\

Despite its success, the exact AGN heating mechanisms are still not fully understood. Various processes have been proposed, including bubble mixing \citep{Hillel_2015, Karen_Yang_2016}, weak shock heating \citep{Fabian_2003,Nulsen_2005}, thermal conduction \citep{Voigt2004, Yang2016a}, turbulence \citep{David_2001,Kim_2003}, and cosmic ray heating \citep{Guo_2008,Pfrommer_2013,Ruszkowski_2017,Yang_2019}. In addition to the relative importance of these mechanisms, some fundamental physical properties of AGN jets, such as jet density, may also play a critical role in determining the overall heating efficiency. Using magnetohydrodynamic (MHD) simulations of self-regulated AGN feedback, \citet{Ehlert_2022} suggested that lighter jets can achieve more efficient and isotropic heating of the ICM.\\

However, the question remains: how does jet density directly affect the heating efficiency and isotropy? In this work, to build upon the findings of \citet{Ehlert_2022} and to isolate the effect of jet density in a more controlled model, we perform three-dimensional (3D) hydrodynamic (HD) simulations with cold-mode accretion. In both single-jet ejection and self-regulated feedback scenarios, we vary the jet density to closely examine its impact on the heating process. Our results show that lighter jets can indeed heat the cluster more isotropically and influence the heating efficiency. This suggests that jet density is a key parameter in realistic AGN feedback modeling.\\

The outline of this paper is as follows, We summarize the essential elements of our simulation setups in Section \ref{sec:methods}. The simulation results of the single-jet and self-regulated setups are presented in Sections \ref{subsec:singlejet} and \ref{subsec:self-regulated}, respectively. We test the convergence of our results in Section \ref{subsec:HR}, compare with previous studies in Section \ref{subsec: discuss-compare}, and discuss the limitations of our simulations in Section \ref{subsec:limit}. Finally, we conclude our findings in Section \ref{sec:conclusions}.

\section{Methods}
\label{sec:methods}

We carry out 3D HD simulations for two different models, single-jet and self-regulated AGN feedback in an isolated Perseus-like cluster using the adaptive-mesh-refinement (AMR) code FLASH \citep{Flash, Dubey08}. More details of the simulation setup can be found in \citet{Yang_2019} for the single-jet model and \citet{Karen_Yang_2016} for the self-regulated model. Here, we provide a brief summary and highlight the key differences.\\

The ICM in the cluster is initialized by empirical fits to the surface brightness of the observed Perseus cluster assuming hydrostatic equilibrium with a static NFW \citep{Navarro_1996} gravitaional potential:
\begin{equation}
    \Phi(r) = -\frac{GM_\text{vir}}{r}\frac{\ln{(1+r/r_s)}}{\ln (1+c) -c/(1+c)},
\end{equation}
where $M_\text{vir}$ is the cluster virial mass, $r_s \equiv r_\text{vir}/c$ is the scale radius, $r_\text{vir}$ is the virial radius, and $c$ is concentration parameter. We set $M_\text{vir} = 8.5 \times 10^{14} M_\odot$, $r_\text{vir} = 2440 \text{ kpc}$, and $c = 6.81$ for both the single-jet and self-regulated models. Since the gas contributes little to gravity, we neglect the self-gravity effect.
In both models, we add radiative cooling computed by the tabulated table of Sutherland \& Dopita \citeyearpar{SD1993} with $1/3$ solar metallicity. 

\subsection{Single-jet setup}
\label{subsec: single setup}

In the single-jet simulations, the computational domain spans 500 kpc on each side and is adaptively refined based on temperature gradients, reaching up to an AMR level of 8, corresponding to a peak resolution of 0.5 kpc. Reflective boundary conditions are applied; however, due to the relatively short simulation time about 100 Myr, the forward shock generated by the jet injection has not reached the boundary, and the reflected waves are not expected to significantly impact the results. 

In this single-jet model, we set a fixed jet power of $5 \times10^{45}$ erg s$^{-1}$, jet duration of 10 Myr, and inject the jet mass, momentum, and energy into a cylindrical nozzle at the cluster center with radius of 2 kpc ($r_{\rm ej}$) and height of 4 kpc ($h_{\rm ej}$). The jets are ejected in a bipolar fashion along the $z$-axis and are assumed to be purely kinetic. Therefore, the density of jets ($\rho_\text{jet}$) is inversely proportional to $v_\text{jet}^3$:
\begin{equation}
    P_\text{jet} = \rho_\text{jet} \cdot v_\text{jet}^3 \cdot \pi r_\text{ej}^2,
\end{equation}
where $P_\text{jet}$ is jet power, and $v_\text{jet}$ is jet velocity. By assigning different jet velocities, we could obtain different jet densities.
Table \ref{tab: jet density} summarizes the adopted value of jet densities for light, fiducial, and heavy cases.

\subsection{Self-regulated setup}
\label{subsec: sr setup}
In the self-regulated simulations, the domain is 500 kpc on a side and adaptively refined up to an AMR level of 7, which corresponds to a resolution element of 1 kpc. For the converge test shown in Section \ref{subsec:HR}, we perform the high resolution (HR) cases with an AMR max level of 8 within the 30 kpc from the cluster center, corresponding to a resolution element of 0.5 kpc. Diode boundary conditions are applied, which only allow outflows but not inflows. \\

The central supermassive black hole (SMBH) is assumed to be fed through cold-mode accretion, accreting cold gas ($T < 5 \times 10^{5}$ K) in the black hole vicinity that falls in on the free-fall timescale due to gravitational attraction of the SMBH. The accretion rate of SMBH is assumed to be $\dot{M}_{\text{BH}} = M_{\text{cold}} /t_{\text{ff}} $, where $M_{\text{cold}}$ is the total amount of cold gas within an accretion radius $r_{\text{accre}} = 2$ kpc and $t_\text{ff} = 4.5$ Myr is the approximate free-fall time at $r_\text{accr}$ for the simulated cluster. In the self-regulated model, as its name suggests, the accretion rate is used to compute the mass, momentum, and energy injection rates of the jets as follows:
\begin{align}
\label{eq:accretion rate}
    \dot{M} &= \eta \dot{M}_{\text{BH}} \nonumber\\
    |\dot{P}| & = \sqrt{2\eta \epsilon} \dot{M}_{\text{BH}} c \nonumber\\
    \dot{E} & = \epsilon \dot{M}_{\text{BH}} c^2,  
\end{align}
where $\epsilon = 0.001$ is feedback efficiency and $\eta$ is mass loading factor. The jets are assumed to be processing along the $z$-axis with a $15^\circ$ angle and 10 Myr period. The feedback is applied to a cylinder with radius of 2.5 kpc and height of 4 kpc at the cluster center. To investigate the influence of jet density, we vary the mass loading factor, $\eta = $1, 0.1, 0.01, and 0.001, such that in each case the jet density differs by approximately one order of magnitude. The jet parameters are summerized in Table \ref{tab: jet density}.
\begin{table*}
    \centering
    \caption{Simulation parameters in the single-jet and self-regulated models, including jet power, jet density, and jet velocity.}
    \label{tab: jet density}
    \begin{tabular}{lllll}
    \hline
        Model & Case or $\eta$ & Jet power (erg s$^{-1}$)& Jet density ($\text{g cm}^{-3}$)& Jet velocity (c) \\ 
        \hline
        Single jet & Light & $5.00 \times10^{45}$ &  $1.95 \times10^{-28}$ & 0.1998  \\ 
        ~ & Fiducial & $5.00 \times10^{45}$ &  $1.56 \times10^{-27}$ & 0.0999  \\ 
        ~ & Heavy & $5.00 \times10^{45}$ &  $4.92 \times10^{-26}$ & 0.0316  \\ 
        \hline
        Self-regulated & 0.001 & $8.85 \times10^{46}$ & $5.56 \times10^{-28}$ &  0.3160 \\ 
        ~ & 0.01 & $3.43 \times10^{46}$& $6.82 \times10^{-27}$ &  0.1000 \\ 
        ~ & 0.1 & $1.49 \times10^{46}$& $9.39 \times10^{-26}$ &  0.0316 \\ 
        ~ & 1 & $8.70 \times10^{45}$ & $1.73 \times10^{-24}$ &  0.0100 \\ 
        \hline
    \end{tabular}
\end{table*}
\section{Results}
\label{sec:results}
\subsection{Single-jet simulations}
\label{subsec:singlejet}
To understand in detail how the jet density affects the evolution of the ICM, we first perform the single-jet simulations as described in Section \ref{subsec: single setup}. In these simulations, the jets are injected only once with a fixed duration, allowing us to isolate and observe the effect of jet density on its evolution without interference from other factors.\\
\begin{figure*}
    \centering
    \includegraphics[width=0.9\linewidth]{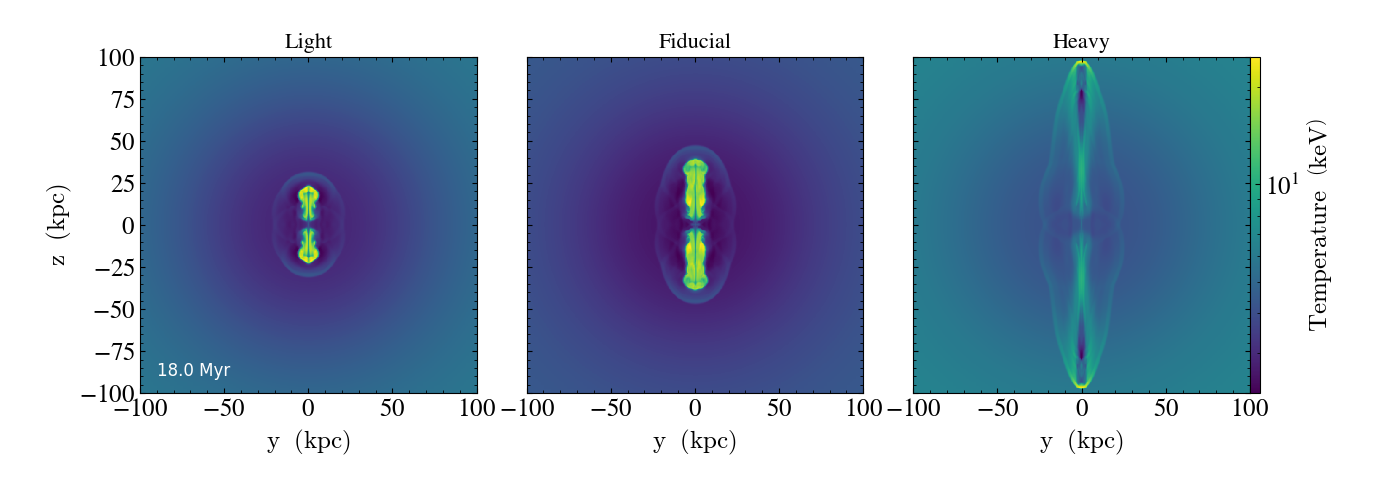}
    \caption{Temperature slice plots at 18 Myr after ejection, comparing different jet density cases in the single-jet model. Lighter jets produce more spherical and hotter bubbles near the cluster center, while heavier jets fail to maintain clear bubbles and penetrate the ICM more quickly.}
    \label{fig:singlejet_tem}
\end{figure*}

In Fig.~\ref{fig:singlejet_tem} we present the temperature slices from the single-jet simulations, which shows the evolution of bubbles for different densities of jets. As the jets start to inject energy, it drives a forward shock that propagates outward from the cluster center, followed by the inflation of a pair of low-density, hot bubbles. Even after the jet activity ceases, the bubbles continue to rise buoyantly through the ICM while expanding, gradually becoming unstable due to hydrodynamic instabilities and mixed with the ambient medium until they are eventually disrupted. \\

Comparing the three cases, we find that the lighter jets tend to remain concentrated near the cluster center for a longer time, producing hotter and more spherical bubbles. This morphology also results in shock waves that are more spherical, promoting efficient heating in the cluster core (see also Section \ref{subsec:self-regulated}). In contrast, the heavier jets, due to their higher momentum, quickly escape the central region and are unable to produce clear and long-lived bubbles.\\

\begin{figure}
    \centering
    \includegraphics[width=0.9\linewidth]{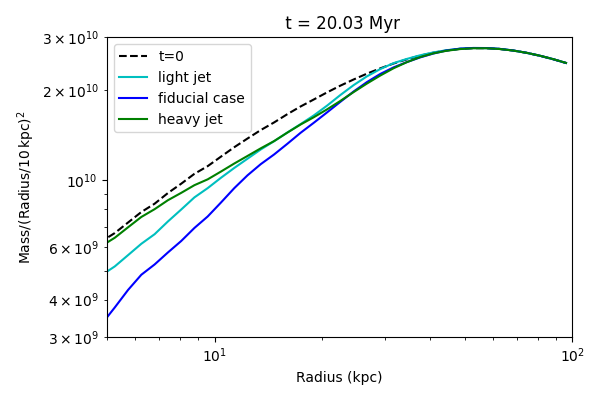}
    \caption{Enclosed mass profiles for single-jet simulations with different jet densities, normalized by radius squared in order to emphasize the differences close to the cluster center. This figure shows the efficiency of the ICM uplifting by jets in difference cases. The black dashed line represents the initial profile at $t = 0$. }
    \label{fig:mass-uplift}
\end{figure}
\begin{figure*}
    \centering
    \includegraphics[width=0.9\linewidth]{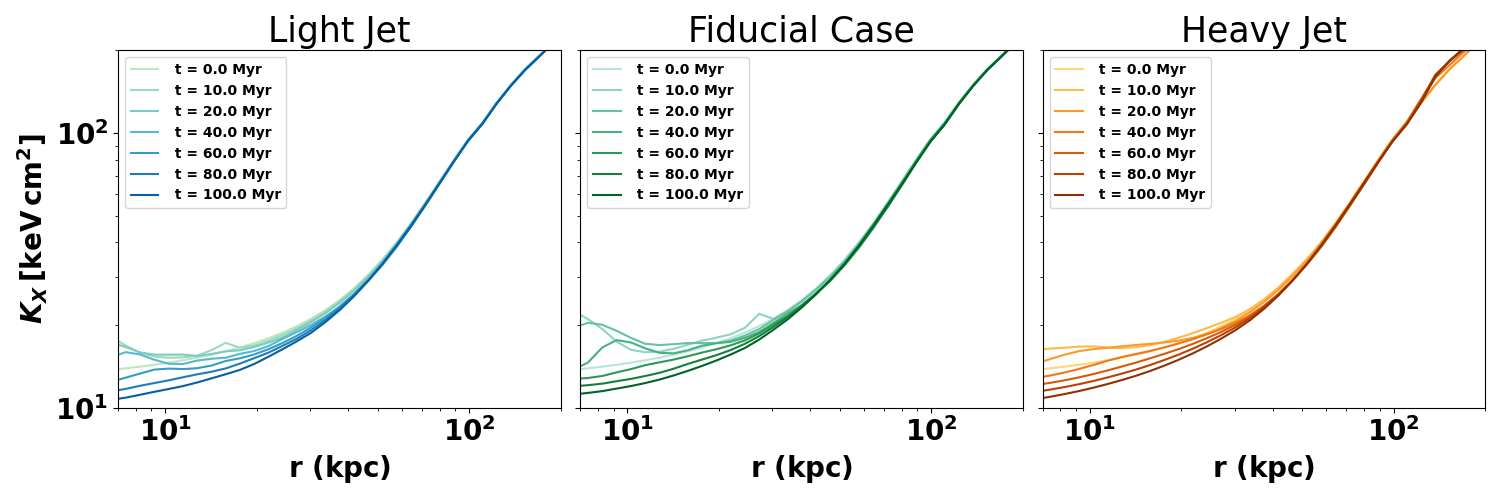}
    \caption{Entropy profiles for the three single-jet simulations: the different colors represent the different simulation times. This figure shows that the fiducial case has the highest heating efficiency in the central region.}
    \label{fig:entropy_profile_singlejet)}
\end{figure*}

Fig.~\ref{fig:mass-uplift} shows the radial mass distributions for the three cases at $t = 20$ Myr. In all three cases, the jets uplift the ICM from the cluster center, which is regarded as a mechanism that helps to alleviate the cooling-flow problem \citep{Karen_Yang_2016,Guo_2017,Zhang_2022,Husko_2023}. The dependence of the uplift efficiency on jet density is not monotonic though, with the fiducial case (blue line) exhibiting the most efficient uplift of the ICM.\\

Fig. \ref{fig:entropy_profile_singlejet)} shows the entropy profiles at different simulation times, where the entropy is defined as $K \equiv k_B T/n_e^{2/3}$. The entropy profiles indicate that the fiducial case produces the strongest heating effect near the cluster center ($r < 20$ kpc). Together with Fig. \ref{fig:mass-uplift}, the above results suggest that the fiducial case is the most effective in heating and displacing the ICM from the center. Notably, this effect exhibits a non-monotonic dependence on jet density---the optimal performance of heating and uplift occurs at an intermediate density, and both lighter and heavier jets are less effective. This may be because of the fact that the heavier jets, with their higher momentum, more easily escape from the cluster center and fail to effectively couple to the ambient ICM, while lighter jets tend to form smaller bubbles near the center. Jets with intermediate densities reach a balance, maintaining bubbles of sufficient sizes to perform effective uplift and heating.

\subsection{Self-regulated simulations}
\label{subsec:self-regulated}
In this section, we investigate how jet density influences the self-regulation process, i.e., the balance between heating and cooling and the overall evolution of the cluster. 
At the beginning of the simulations, the cluster continues to contract due to radiative cooling without any jet activity. After about 300 Myr, jets are triggered: as the fast jets drive into the ICM, they form a bow shock, followed by the inflation of hot expanding bubbles behind the shock front. When the jets temporarily cease, the buoyant bubbles rise through the ICM, inducing hydrodynamic instabilities that promote mixing between the ICM and the bubbles until they are eventually disrupted \citep{Bourne_2023}. This process represents how the kinetic energy of the jets is converted into thermal energy of the ICM. This heating suppresses the formation of cold clumps until the ICM cools again and forms cold clumps that accrete onto the SMBH, which then triggers the next cycle of jet activity. This is how a self-regulated AGN feedback cycle is established \citep{Karen_Yang_2016}. All cases with different mass loading factors reach self-regulation with quasi-steady jet power at around 500 Myr and continue to maintain this state afterwards.\\
\begin{figure*}
    \centering
    \includegraphics[width=0.9\linewidth]{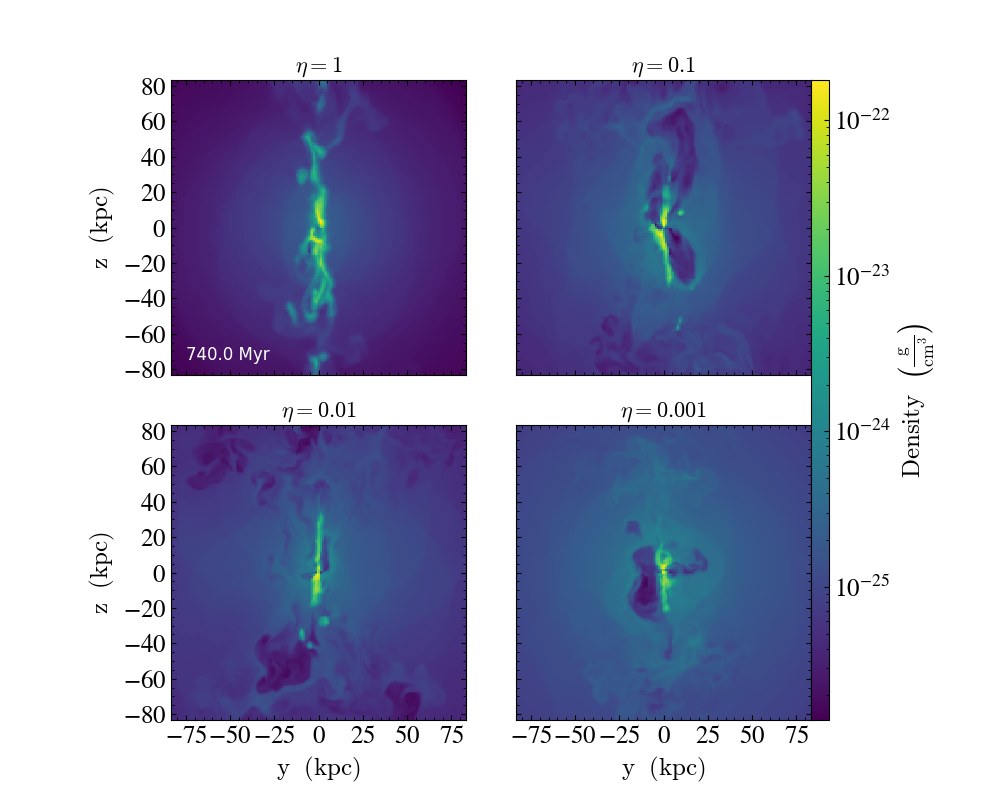}
    \caption{Density slices at $t = $740 Myrs for the self-regulated cases with varied mass loading factors $\eta$ (see Table \ref{tab: jet density}). The bubbles formed by lighter jets have a larger angle of deviation from the jet precession axis than those formed by heavier jets.}
    \label{fig:density_slice_sr}
\end{figure*}

Fig.~\ref{fig:density_slice_sr} shows the density slices at $x = 0$ comparing different mass loading factors $\eta$, in which we can observe the bubble shapes and their propagation directions. The lighter jets ($\eta = 0.001$) produce more spherical bubbles concentrated near the center and exhibit larger angles of deviation from the jet precession axis ($z$-axis). In contrast, the heavier jets ($\eta = 1$ and $0.1$) tend to produce bubbles along the $z$-axis. These features are consistent with the results from the single-jet simulations discussed in \ref{subsec:singlejet}.\\

\begin{figure}
    \centering
    \includegraphics[width=1\linewidth]{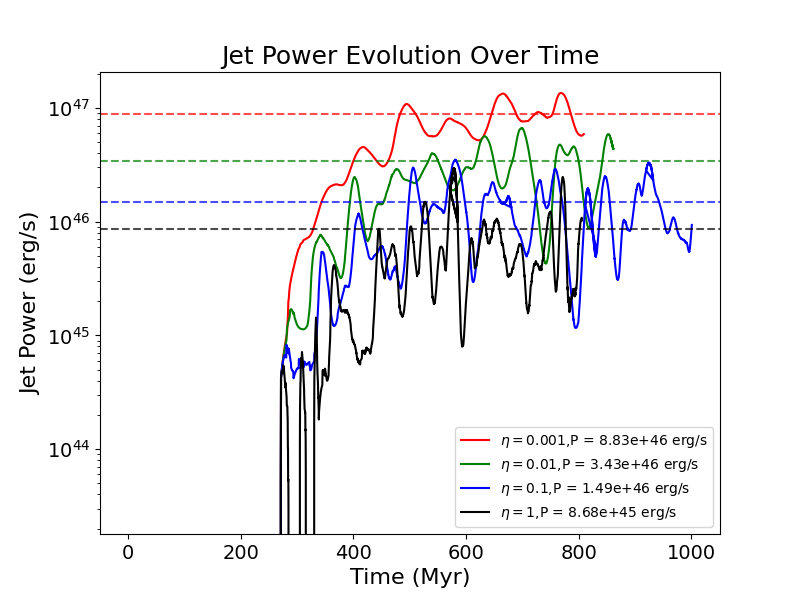}
    \caption{Evolution of jet power for the self-regulated simulations with varied mass loading factors. When $t \gtrsim 500$ Myr, the cluster enters a self-regulated state, as the jet power approaches a quasi-equilibrium value. The time-averaged jet power after $t = 500$ Myr is plotted using the horizontal dashed lines and is indicated in the legend for each case. This figure shows that the lighter jets have higher time-averaged jet power and thus lower heating efficiency compared to heavier jets.}
    \label{fig:powerjet_nohr}
\end{figure}

To more clearly quantify the heating efficiency of each case, we compare the jet power required to heat the CC cluster and maintain its self-regulated state. Fig.~\ref{fig:powerjet_nohr} shows the evolution of AGN jet power and its average value for simulations with different mass loading factors. Our simulations indicate that the lighter jets exhibit higher time-averaged jet power, which implies lower ICM heating efficiencies. This may be because the lighter jets have less momentum than heavier jets to disperse the cold gas, and hence more jet power is wasted to heat cold clumps with excessively short cooling times. This trend has a monotonic dependence on jet density, and interestingly, is opposite to the finding of \citet{Ehlert_2022} who found that lighter jets tend to have better heating efficiencies. We discuss possible reasons for this difference in Section~\ref{sec:discussions}.\\
\begin{figure*}
    \centering
    \includegraphics[width=0.85\linewidth]{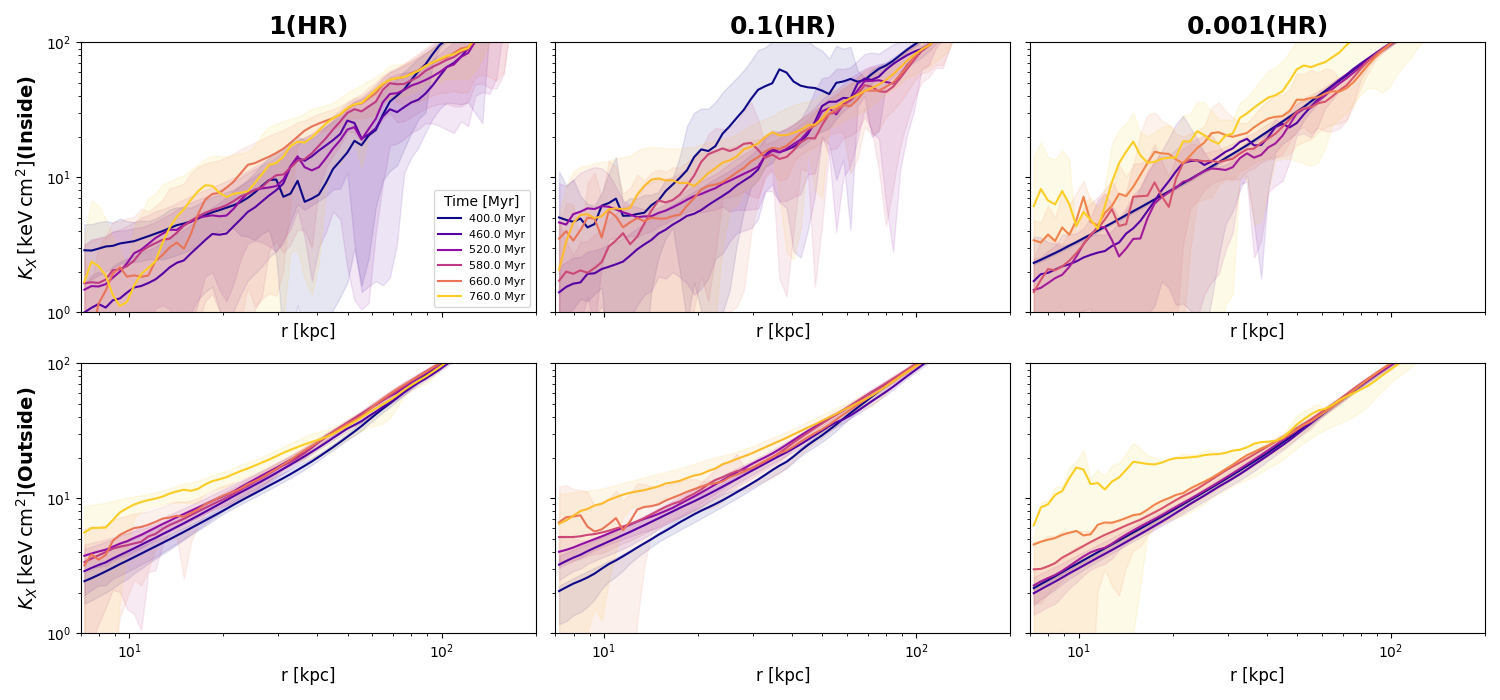}
    \caption{Entropy profiles for the high-resolution, self-regulated simulations with varied mass loading factors inside (top row) and outside (bottom row) the jet cones, defined as the region within $30^{\circ}$ with respect to the jet precession axis. The shaded areas represent the 10th to 90th percentile ranges over the time intervals between each solid line. It is clear that the $\eta = 0.001$ case has much more effective heating to the region outside the jet cones, indicating that the lighter jets could heat the ICM more isotropically.}
    \label{fig:entropy_jetcone}
\end{figure*}

 To further investigate the spatial distribution and isotropy of jet heating, we divide the simulation domain into regions inside and outside the jet cones, defined as the region within $30^\circ$ with respect to the jet precession axis. Fig. \ref{fig:entropy_jetcone} shows the entropy profiles from the HR simulations at different times, separated into regions inside and outside the jet cones.\footnote{We show the HR simulations here because the trends in the fiducial simulations, though consistent, is less apparent compared to the HR cases.} Within the jet cones, all cases exhibit similar heating behavior with large fluctuations, primarily due to the intermittent nature of the jet outbursts. However, outside the jet cones, the lighter jets ($\eta = 0.001$) demonstrate more significant heating compared to the heavier jets ($\eta = $1 and 0.1), indicating a more isotropic heating effect of the lighter jets. This result is consistent with the more spherical and isotropic bubble morphology observed in the density slices shown in Fig. \ref{fig:density_slice_sr} as well as the findings of \citet{Ehlert_2022}.\\

\section{Discussion}
\label{sec:discussions}
\subsection{High-resolution simulations}
\label{subsec:HR}

To verify the convergence of our simulation results and investigate the discrepancy with results in the previous work, we perform additional HR simulations with self-regulated AGN feedback. The setup of the HR simulation is mentioned in Section \ref{subsec: sr setup}. The simulations of higher resolution will provide finer spatial details, such as the structures of the cold gas, but the overall cluster evolution in the HR simulations are largely consistent with the fiducial runs.\\
\begin{figure*}
    \centering
    \includegraphics[width=1\linewidth]{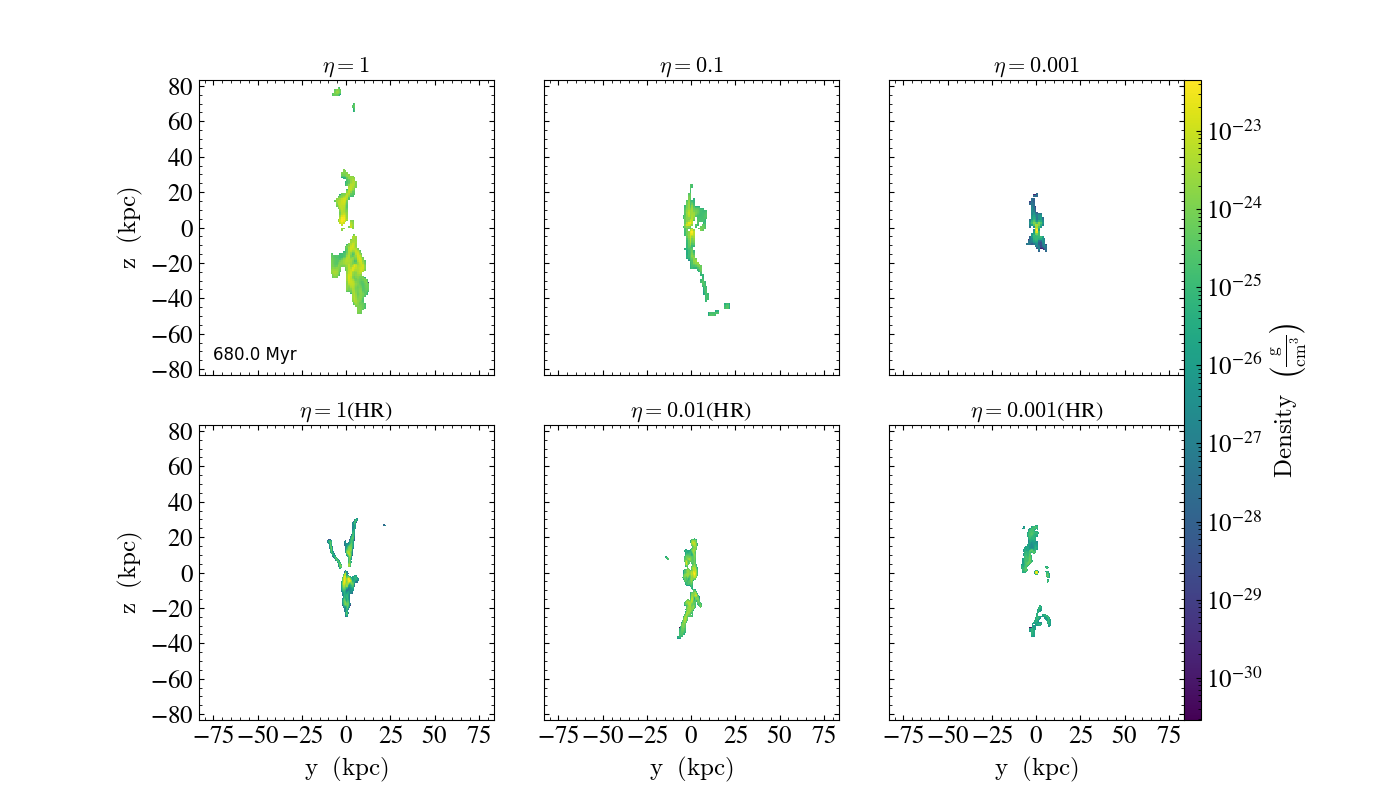}
    \caption{Cold gas distributions of the fiducial (top row) and HR cases (bottom row). In general, we find less accumulation of cold gas near the cluster center in the HR cases.}
    \label{fig:coldgas(HR)}
\end{figure*}

Fig.~\ref{fig:coldgas(HR)} shows the density slices of the cold gas in the fiducial and HR cases. It is evident that the amount of cold gas decreases and becomes more fragmented in the HR runs. The central cold gas density is also lower, indicating that there is less accumulation of cold gas near the cluster center in the HR simulations. This may help reduce the excessive and inefficient heating of the dense cold gas as discussed in Section \ref{subsec:self-regulated}.\\
\begin{figure}
    \centering
    \includegraphics[width=1\linewidth]{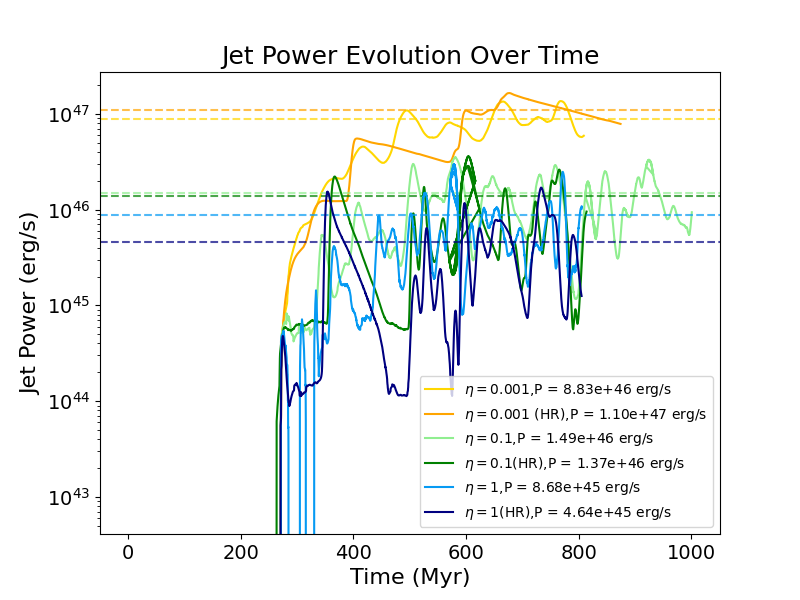}
    \caption{Evolution of jet power for different mass loading factors comparing the fiducial and HR simulations, showing consistent results on the average jet power between the fiducial and HR runs.}
    \label{fig:powerjet(HR)}
\end{figure}

Fig.~\ref{fig:powerjet(HR)} is similar to Fig. \ref{fig:powerjet_nohr}, showing the jet power comparing the fiducial and HR cases. Overall, the HR results are consistent with those of the fiducial cases, showing that the averaged jet power is higher for simulations with lower jet densities. For simulations with $\eta =$ 0.1 and 0.001, the average jet power remains nearly the same---around $10^{47}\ \text{erg s}^{-1}$ and $1.4 \times 10^{46}\ \text{erg s}^{-1}$, respectively---demonstrating good convergence in the HR simulations. However, for the $\eta = $ 1.0 case, the average power in the HR simulation is significantly reduced by nearly a factor of two. We inspect the total amount of cold gas in these simulations and found that the difference in jet power is likely due to the substantial reduction in cold gas mass in the HR run of the $\eta = $1.0 case. In contrast, the simulations with $\eta = 0.1$ and 0.001 contain relatively little cold gas, and the HR primarily alters its distribution without significantly changing the total amount. 

These results suggest that the amount of cold gas can significantly influence the total power required to maintain self-regulation, while its distribution mainly impacts how isotropic the heating is distributed. Overall, we have confirmed the convergence of the relationship between jet power and jet density, as the same trend persists in the HR cases.

\subsection{Comparison with previous works}
\label{subsec: discuss-compare}
While our results confirm the finding in \citet{Ehlert_2022} that lighter jets tend to heat the ICM more isotropically, it appears that our results differ in some aspects from those of \citet{Ehlert_2022}, particularly regarding the dependence of the heating efficiency on jet densities. In this section, we further compare the differences in our simulation setups and discuss possible reasons that may account for the difference in the simulation outcomes.\\

Specifically, we compare our cases with those in \citet{Ehlert_2022} regarding the order of magnitude of jet density. Our self-regulated simulation with $\eta =$ 0.001 corresponds approximately to their "Fiducial" case, with a jet density of $\sim10^{-28}\ \text{g cm}^{-3}$, while our $\eta = 0.1$ case aligns with their "Dense" case at $\sim10^{-25}\ \text{g cm}^{-3}$. In contrast to their study, we introduce a finer sampling of jet densities to more precisely examine how jet density affects the heating efficiency.\\

Another key difference is that they performed simulations with different jet densities using magnetohydrodynamics (MHD) simulations, in contrast to our HD simulations. One key result from their study is that jets can amplify the local magnetic field, potentially leading to strong magnetic braking of cold gas velocities \citep{Wang_2020}. The amplified field can be stretched by the flow and exert a tension force, which transforms cold gas into elongated filaments and suppresses the formation of central cold disks. This is consistent with our discussion in the previous sections, where we noted that lighter jets tend to waste energy when attempting to heat dense cold gas near the cluster center. In the absence of a cold disk, less jet power may be required to achieve effective heating. Moreover, \citet{Ehlert_2022} pointed out that lighter jets may amplify magnetic fields more isotropically. This implies that jets with different densities could interact with cold gas undergoing different levels of magnetic braking, potentially explaining the opposite trend we observe in the relationship between jet power and jet density.\\

We also note that there are significant differences in the absolute values of several key properties. In some of our simulations, the total cold gas mass could reach approximately $10^{12}~M_\odot$, which is considerably higher than the $\sim10^{9}~M_\odot$ cold gas mass reported by \citet{Ehlert_2022}—and the spatial distribution of cold gas also differs markedly. This discrepancy contributes to the higher average jet power in some of our simulations ($10^{46} - 10^{47}$ erg s$^{-1}$), compared to the $\sim10^{45}$ erg s$^{-1}$ reported in their work. These differences could partly explain the discrepancies observed between our results and those of \citet{Ehlert_2022}. When compared with observational results of the Perseus cluster, a more realistic range for the cold gas mass and jet power would be on the order of $10^{9} - 10^{10}~M_\odot$ and $10^{43} - 10^{46}$ erg s$^{-1}$ \citep{PETERSON20061}, respectively. In our simulations, the jet power in the $\eta =$ 0.001 case and the cold gas mass in the $\eta =$ 1.0 case both exceed observational values by more than an order of magnitude, suggesting that these cases may not accurately represent the Perseus cluster. This also implies that in realistic cases, the jet densities likely lie in the intermediate ranges. Nonetheless, since our primary goal is to isolate and understand the impact of jet density on heating efficiency, our setup still provides a clean and direct comparison, largely free from other factors.

\subsection{Limitations of our simulations}
\label{subsec:limit}
In this study, we performed HD simulations and focused specifically on the heating effects driven by bubble expansion induced by jets with different densities. In order to isolate the impact of jet density on feedback, we intentionally neglected other factors, including magnetic field, viscosity, and cosmic rays. We further discuss the possible impacts of these physical mechanisms.\\

In a viscous medium, viscosity can help preserve the AGN jet-inflated bubbles, preventing their disruption over longer timescales due to hydrodynamic instabilities \citep{Reynolds_2005,Dong_2009,Kingsland_2019}. Observational evidence for this effect includes the smooth, coherent bubble shapes seen in the Perseus cluster \citep{Fabian_2003} and long, coherent $H_\alpha$ filaments in the wakes of the uplifted bubbles, suggesting laminar flows \citep{Reynolds_2005}. Viscosity also facilitates the rapid dissipation of sound waves generated by bubble expansion, promoting more isotropic energy transfer throughout the ICM \citep{Ruszkowski_2004}. When thermal conduction is also present, this form of heating can become dominant over radiative cooling above a certain temperature threshold \citep{Fabian_2005}. These physical processes are likely to enhance the overall heating efficiency of AGN feedback \citep{Wang_2022}.\\

Other heating mechanisms may also play important roles. Cosmic rays can provide additional pressure support to AGN-inflated bubbles, enabling them to uplift gas more effectively and thereby altering the distribution and amount of cold gas \citep{Yang_2019}, which could influence the heating efficiency observed in our simulation results. Turbulence arising from various sources can also contribute to heating by displacing rising bubbles and promoting a more isotropic distribution of energy \citep{Morsony_2010}.

\section{Conclusions}
\label{sec:conclusions}
To address the cooling-flow problem in CC clusters, it is essential to understand the overall picture of AGN feedback and its heating mechanisms. In this study, we conducted HD simulations of a Perseus-like cluster, including both single-jet injections and self-regulated AGN feedback, with a focus on how varying jet densities influences the heating efficiency and spatial distribution of energy deposition. We found that the heating performance indeed differs with jet density. Specifically, our results show that:

\begin{enumerate}
    \item Lighter jets are capable of heating the ICM more isotropically. They tend to produce more spherical bubbles and are more easily deflected by cold gas clumps, allowing their directions to change more readily. As a result, lighter jets can distribute energy not only along the jet precession axis, but also more broadly throughout the cluster core. This result is consistent with the finding of \citet{Ehlert_2022}
    \item Although lighter jets tend to heat the ICM more isotropically, our simulations show that they exhibit lower overall heating efficiencies compared to heavier jets. This is reflected in the higher average jet power required to maintain a self-regulated state, indicating that more jet energy is needed to achieve the thermal balance with radiative cooling. This trend contrasts with previous studies  \citep{Ehlert_2022} and the possible reasons are discussed in Section 4.2.
    \item The spatial distribution of cold gas could impact the isotropy of heating, while the amount of cold gas influences the heating efficiency. A high concentration of dense cold gas near the cluster center would lead to inefficient heating of the low-density jets. 
\end{enumerate}
Our current study underscores the role of jet density in determining the dynamical and thermal evolution of the ICM. As discussed in Section~\ref{sec:discussions}, various other physical processes—such as magnetic fields, viscosity, and cosmic rays—can further alter the evolution, distribution, and total mass of cold gas, thereby affecting the efficiencies and distributions of heating by AGN jets. Future studies incorporating the above mechanisms would be needed in order to make comparisons with observations and further constrain realistic ranges of the jet densities. 

\section*{Acknowledgements}

TWT acknowledges the National Center of Theoretical Sciencies (NCTS) Summer Student Program 2024 for providing the opportunity to begin this research and thanks HYKY for valuable guidance and support throughout this work. This study made use of the FLASH code and the yt analysis tool, as well as the high-performance computing facilities operated by Center for Informatics and Computational in Astronomy (CICA) at National Tsing Hua University (NTHU). HYKY acknowledges support from the National Science and Technology Council (NSTC) of Taiwan (NSTC 112-2628-M-007-003-MY3; NSTC 114-2112-M-007-032-MY3) and the Yushan Scholar Program of the Ministry of Education (MoE) of Taiwan (MOE-108-YSFMS-0002-003-P1).

\section*{Data Availability}

The simulations were performed using the code FLASH, publicly available at \url{https://flash.rochester.edu/site/flashcode/}. Source data are available from the corresponding author upon reasonable request.



\bibliographystyle{mnras}
\bibliography{mnras} 

\begin{thebibliography}{}
\makeatletter
\relax
\def\mn@urlcharsother{\let\do\@makeother \do\$\do\&\do\#\do\^\do\_\do\%\do\~}
\def\mn@doi{\begingroup\mn@urlcharsother \@ifnextchar [ {\mn@doi@} {\mn@doi@[]}}
\def\mn@doi@[#1]#2{\def\@tempa{#1}\ifx\@tempa\@empty \href {http://dx.doi.org/#2} {doi:#2}\else \href {http://dx.doi.org/#2} {#1}\fi \endgroup}
\def\mn@eprint#1#2{\mn@eprint@#1:#2::\@nil}
\def\mn@eprint@arXiv#1{\href {http://arxiv.org/abs/#1} {{\tt arXiv:#1}}}
\def\mn@eprint@dblp#1{\href {http://dblp.uni-trier.de/rec/bibtex/#1.xml} {dblp:#1}}
\def\mn@eprint@#1:#2:#3:#4\@nil{\def\@tempa {#1}\def\@tempb {#2}\def\@tempc {#3}\ifx \@tempc \@empty \let \@tempc \@tempb \let \@tempb \@tempa \fi \ifx \@tempb \@empty \def\@tempb {arXiv}\fi \@ifundefined {mn@eprint@\@tempb}{\@tempb:\@tempc}{\expandafter \expandafter \csname mn@eprint@\@tempb\endcsname \expandafter{\@tempc}}}

\bibitem[\protect\citeauthoryear{{Beckmann, R. S.} et~al.,}{{Beckmann, R. S.} et~al.}{2019}]{Beckmann_2019}
{Beckmann, R. S.} et~al., 2019, \mn@doi [A&A] {10.1051/0004-6361/201936188}, 631, A60

\bibitem[\protect\citeauthoryear{Bourne \& Yang}{Bourne \& Yang}{2023}]{Bourne_2023}
Bourne M.~A.,  Yang H.-Y.~K.,  2023, \mn@doi [Galaxies] {10.3390/galaxies11030073}, 11, 73

\bibitem[\protect\citeauthoryear{David, Nulsen, McNamara, Forman, Jones, Ponman, Robertson  \& Wise}{David et~al.}{2001}]{David_2001}
David L.~P.,  Nulsen P. E.~J.,  McNamara B.~R.,  Forman W.,  Jones C.,  Ponman T.,  Robertson B.,   Wise M.,  2001, \mn@doi [The Astrophysical Journal] {10.1086/322250}, 557, 546

\bibitem[\protect\citeauthoryear{Dong \& Stone}{Dong \& Stone}{2009}]{Dong_2009}
Dong R.,  Stone J.~M.,  2009, \mn@doi [The Astrophysical Journal] {10.1088/0004-637X/704/2/1309}, 704, 1309

\bibitem[\protect\citeauthoryear{{Dubey}, {Reid}  \& {Fisher}}{{Dubey} et~al.}{2008}]{Dubey08}
{Dubey} A.,  {Reid} L.~B.,   {Fisher} R.,  2008, Physica Scripta, T132, p. 014046

\bibitem[\protect\citeauthoryear{Dunn \& Fabian}{Dunn \& Fabian}{2008}]{Dunn_2008}
Dunn R. J.~H.,  Fabian A.~C.,  2008, Monthly Notices of the Royal Astronomical Society, 385, 757

\bibitem[\protect\citeauthoryear{Ehlert, Weinberger, Pfrommer, Pakmor  \& Springel}{Ehlert et~al.}{2022}]{Ehlert_2022}
Ehlert K.,  Weinberger R.,  Pfrommer C.,  Pakmor R.,   Springel V.,  2022, \mn@doi [Monthly Notices of the Royal Astronomical Society] {10.1093/mnras/stac2860}, 518, 4622–4645

\bibitem[\protect\citeauthoryear{Fabian}{Fabian}{1994}]{Fabian_1994}
Fabian A.~C.,  1994, \mn@doi [Annual Review of Astronomy and Astrophysics] {https://doi.org/10.1146/annurev.aa.32.090194.001425}, 32, 277

\bibitem[\protect\citeauthoryear{Fabian}{Fabian}{2012}]{Fabian_2012}
Fabian A.,  2012, \mn@doi [Annual Review of Astronomy and Astrophysics] {https://doi.org/10.1146/annurev-astro-081811-125521}, 50, 455

\bibitem[\protect\citeauthoryear{Fabian, Sanders, Allen, Crawford, Iwasawa, Johnstone, Schmidt  \& Taylor}{Fabian et~al.}{2003}]{Fabian_2003}
Fabian A.~C.,  Sanders J.~S.,  Allen S.~W.,  Crawford C.~S.,  Iwasawa K.,  Johnstone R.~M.,  Schmidt R.~W.,   Taylor G.~B.,  2003, \mn@doi [Monthly Notices of the Royal Astronomical Society] {10.1046/j.1365-8711.2003.06902.x}, 344, L43–L47

\bibitem[\protect\citeauthoryear{Fabian, Reynolds, Taylor  \& Dunn}{Fabian et~al.}{2005}]{Fabian_2005}
Fabian A.~C.,  Reynolds C.~S.,  Taylor G.~B.,   Dunn R. J.~H.,  2005, \mn@doi [Monthly Notices of the Royal Astronomical Society] {10.1111/j.1365-2966.2005.09484.x}, 363, 891–896

\bibitem[\protect\citeauthoryear{{Fryxell}, {Olson}, {Ricker}  et~al.}{{Fryxell} et~al.}{2000}]{Flash}
{Fryxell} B.,  {Olson} K.,  {Ricker} P.,   et~al., 2000, \mn@doi [\apjs] {10.1086/317361}, \href {http://adsabs.harvard.edu/abs/2000ApJS..131..273F} {131, 273}

\bibitem[\protect\citeauthoryear{Gaspari, Melioli, Brighenti  \& D'Ercole}{Gaspari et~al.}{2011}]{Gaspari_2011}
Gaspari M.,  Melioli C.,  Brighenti F.,   D'Ercole A.,  2011, \mn@doi [Monthly Notices of the Royal Astronomical Society] {10.1111/j.1365-2966.2010.17688.x}, 411, 349

\bibitem[\protect\citeauthoryear{Gaspari, Ruszkowski  \& Sharma}{Gaspari et~al.}{2012}]{Gaspari_2012}
Gaspari M.,  Ruszkowski M.,   Sharma P.,  2012, \mn@doi [The Astrophysical Journal] {10.1088/0004-637X/746/1/94}, 746, 94

\bibitem[\protect\citeauthoryear{Guo \& Oh}{Guo \& Oh}{2008}]{Guo_2008}
Guo F.,  Oh S.~P.,  2008, Monthly Notices of the Royal Astronomical Society, 384, 251

\bibitem[\protect\citeauthoryear{Guo, Duan  \& Yuan}{Guo et~al.}{2017}]{Guo_2017}
Guo F.,  Duan X.,   Yuan Y.-F.,  2017, \mn@doi [Monthly Notices of the Royal Astronomical Society] {10.1093/mnras/stx2404}, 473, 1332–1345

\bibitem[\protect\citeauthoryear{Hillel \& Soker}{Hillel \& Soker}{2015}]{Hillel_2015}
Hillel S.,  Soker N.,  2015, Monthly Notices of the Royal Astronomical Society, 455, 2139

\bibitem[\protect\citeauthoryear{Huško \& Lacey}{Huško \& Lacey}{2023}]{Husko_2023}
Huško F.,  Lacey C.~G.,  2023, Monthly Notices of the Royal Astronomical Society, 521, 4375

\bibitem[\protect\citeauthoryear{Kim \& Narayan}{Kim \& Narayan}{2003}]{Kim_2003}
Kim W.-T.,  Narayan R.,  2003, \mn@doi [The Astrophysical Journal] {10.1086/379342}, 596, L139

\bibitem[\protect\citeauthoryear{Kingsland, Yang, Reynolds  \& Zuhone}{Kingsland et~al.}{2019}]{Kingsland_2019}
Kingsland M.,  Yang H.-Y.~K.,  Reynolds C.~S.,   Zuhone J.~A.,  2019, \mn@doi [The Astrophysical Journal Letters] {10.3847/2041-8213/ab40be}, 883, L23

\bibitem[\protect\citeauthoryear{Li, Bryan, Ruszkowski, Voit, O’Shea  \& Donahue}{Li et~al.}{2015}]{Li_2015}
Li Y.,  Bryan G.~L.,  Ruszkowski M.,  Voit G.~M.,  O’Shea B.~W.,   Donahue M.,  2015, \mn@doi [The Astrophysical Journal] {10.1088/0004-637x/811/2/73}, 811, 73

\bibitem[\protect\citeauthoryear{Li, Ruszkowski  \& Bryan}{Li et~al.}{2017}]{Li_2017}
Li Y.,  Ruszkowski M.,   Bryan G.~L.,  2017, \mn@doi [The Astrophysical Journal] {10.3847/1538-4357/aa88c1}, 847, 106

\bibitem[\protect\citeauthoryear{McNamara \& Nulsen}{McNamara \& Nulsen}{2012}]{McNamara_2012}
McNamara B.~R.,  Nulsen P. E.~J.,  2012, \mn@doi [New Journal of Physics] {10.1088/1367-2630/14/5/055023}, 14, 055023

\bibitem[\protect\citeauthoryear{Morsony, Heinz, Brüggen  \& Ruszkowski}{Morsony et~al.}{2010}]{Morsony_2010}
Morsony B.~J.,  Heinz S.,  Brüggen M.,   Ruszkowski M.,  2010, \mn@doi [Monthly Notices of the Royal Astronomical Society] {10.1111/j.1365-2966.2010.17059.x}, 407, 1277–1289

\bibitem[\protect\citeauthoryear{Navarro, Frenk  \& White}{Navarro et~al.}{1996}]{Navarro_1996}
Navarro J.~F.,  Frenk C.~S.,   White S. D.~M.,  1996, \mn@doi [The Astrophysical Journal] {10.1086/177173}, 462, 563

\bibitem[\protect\citeauthoryear{Nulsen, McNamara, Wise  \& David}{Nulsen et~al.}{2005}]{Nulsen_2005}
Nulsen P. E.~J.,  McNamara B.~R.,  Wise M.~W.,   David L.~P.,  2005, \mn@doi [The Astrophysical Journal] {10.1086/430845}, 628, 629

\bibitem[\protect\citeauthoryear{Peterson \& Fabian}{Peterson \& Fabian}{2006}]{PETERSON20061}
Peterson J.,  Fabian A.,  2006, \mn@doi [Physics Reports] {https://doi.org/10.1016/j.physrep.2005.12.007}, 427, 1

\bibitem[\protect\citeauthoryear{Pfrommer}{Pfrommer}{2013}]{Pfrommer_2013}
Pfrommer C.,  2013, \mn@doi [The Astrophysical Journal] {10.1088/0004-637X/779/1/10}, 779, 10

\bibitem[\protect\citeauthoryear{Prasad, Sharma  \& Babul}{Prasad et~al.}{2015}]{Prasad_2015}
Prasad D.,  Sharma P.,   Babul A.,  2015, \mn@doi [The Astrophysical Journal] {10.1088/0004-637X/811/2/108}, 811, 108

\bibitem[\protect\citeauthoryear{Reynolds, McKernan, Fabian, Stone  \& Vernaleo}{Reynolds et~al.}{2005}]{Reynolds_2005}
Reynolds C.~S.,  McKernan B.,  Fabian A.~C.,  Stone J.~M.,   Vernaleo J.~C.,  2005, Monthly Notices of the Royal Astronomical Society, 357, 242

\bibitem[\protect\citeauthoryear{Ruszkowski, Bruggen  \& Begelman}{Ruszkowski et~al.}{2004}]{Ruszkowski_2004}
Ruszkowski M.,  Bruggen M.,   Begelman M.~C.,  2004, \mn@doi [The Astrophysical Journal] {10.1086/422158}, 611, 158–163

\bibitem[\protect\citeauthoryear{{Ruszkowski}, {Yang}  \& {Reynolds}}{{Ruszkowski} et~al.}{2017}]{Ruszkowski_2017}
{Ruszkowski} M.,  {Yang} H. Y.~K.,   {Reynolds} C.~S.,  2017, \mn@doi [\apj] {10.3847/1538-4357/aa79f8}, \href {https://ui.adsabs.harvard.edu/abs/2017ApJ...844...13R} {844, 13}

\bibitem[\protect\citeauthoryear{{Sutherland} \& {Dopita}}{{Sutherland} \& {Dopita}}{1993}]{SD1993}
{Sutherland} R.~S.,  {Dopita} M.~A.,  1993, \mn@doi [\apjs] {10.1086/191823}, \href {https://ui.adsabs.harvard.edu/abs/1993ApJS...88..253S} {88, 253}

\bibitem[\protect\citeauthoryear{{Voigt} \& {Fabian}}{{Voigt} \& {Fabian}}{2004}]{Voigt2004}
{Voigt} L.~M.,  {Fabian} A.~C.,  2004, \mn@doi [\mnras] {10.1111/j.1365-2966.2004.07285.x}, \href {http://adsabs.harvard.edu/abs/2004MNRAS.347.1130V} {347, 1130}

\bibitem[\protect\citeauthoryear{Wang \& Yang}{Wang \& Yang}{2022}]{Wang_2022}
Wang S.-C.,  Yang H.-Y.~K.,  2022, \mn@doi [Monthly Notices of the Royal Astronomical Society] {10.1093/mnras/stac788}, 512, 5100–5109

\bibitem[\protect\citeauthoryear{Wang, Ruszkowski  \& Yang}{Wang et~al.}{2020}]{Wang_2020}
Wang C.,  Ruszkowski M.,   Yang H.-Y.~K.,  2020, \mn@doi [Monthly Notices of the Royal Astronomical Society] {10.1093/mnras/staa550}, 493, 4065–4076

\bibitem[\protect\citeauthoryear{{Yang} \& {Reynolds}}{{Yang} \& {Reynolds}}{2016a}]{Yang2016a}
{Yang} H.-Y.~K.,  {Reynolds} C.~S.,  2016a, \mn@doi [\apj] {10.3847/0004-637X/818/2/181}, \href {http://adsabs.harvard.edu/abs/2016ApJ...818..181Y} {818, 181}

\bibitem[\protect\citeauthoryear{Yang \& Reynolds}{Yang \& Reynolds}{2016b}]{Karen_Yang_2016}
Yang H.-Y.~K.,  Reynolds C.~S.,  2016b, \mn@doi [The Astrophysical Journal] {10.3847/0004-637x/829/2/90}, 829, 90

\bibitem[\protect\citeauthoryear{Yang, Gaspari  \& Marlow}{Yang et~al.}{2019}]{Yang_2019}
Yang H.-Y.~K.,  Gaspari M.,   Marlow C.,  2019, \mn@doi [The Astrophysical Journal] {10.3847/1538-4357/aaf4bd}, 871, 6

\bibitem[\protect\citeauthoryear{Zhang, Zhuravleva, Gendron-Marsolais, Churazov, Schekochihin  \& Forman}{Zhang et~al.}{2022}]{Zhang_2022}
Zhang C.,  Zhuravleva I.,  Gendron-Marsolais M.-L.,  Churazov E.,  Schekochihin A.~A.,   Forman W.~R.,  2022, \mn@doi [Monthly Notices of the Royal Astronomical Society] {10.1093/mnras/stac2282}, 517, 616–631

\makeatother
\end{thebibliography}








\bsp	
\label{lastpage}
\end{document}